\documentclass{aa}  
\usepackage{gensymb}
\usepackage{graphicx}
\usepackage{txfonts}
\usepackage[colorlinks,citecolor=blue,urlcolor=blue,filecolor=blue,linkcolor=blue]{hyperref}
\usepackage{xcolor}
\usepackage[colorlinks,citecolor=blue,urlcolor=blue,filecolor=blue,linkcolor=blue]{hyperref}

\newcommand{\B}[1]{#1}  % changes are not marked

\begin{document}

   \title{Contamination of spectroscopic observations by satellite constellations
   %\thanks{\R{thanks}}
   }

   %\subtitle{subtitle}

   \author{O.~R. Hainaut
          \inst{1}
          \and
          S. Moehler\inst{1}
          }

   \institute{European Southern Observatory,
        Karl-Schwarzschild-Stra\ss e, 2
        85470 Garching-bei-M\"unchen,
        Germany\\
        \email{ohainaut@eso.org}
        }

   \date{Received -- ; accepted --}

% \abstract{}{}{}{}{} 
% 5 {} token are mandatory
 
  \abstract
{With the onset of large telecommunication constellations, the number of satellites on low orbit has dramatically increased over the past years, raising concerns among the astronomical community about their impact on observations.}
{Because spectroscopic observations represent a large fraction of professional astronomical observations, and because spectrographs lack spatial information that can reveal the presence of a satellite, this paper focuses on how satellites affect spectroscopic optical observations.
}
{We simulated how often satellites contaminate spectrograph observations by using realistic constellations with over 400,000 objects. We counted the satellites that crossed a representative $10\times 1''$ slit during a 1000\,s exposure, for different solar elevations and observation directions.\\
We also measured how a satellite affects a spectrum by using real data from different scientific targets and a scaled solar analogue as the satellite. We then used standard tools to measure astrophysical parameters and compare them with the clean spectrum.}
{As expected, the fraction of affected spectra varies dramatically with the direction of observation and the elevation of the sun, with a maximum of 10\% at twilight down to no contamination when the sun reaches $-30\degree$, with a nightly average of $\sim2$\%. The probability of a satellite crossing over the astronomical target will be further reduced depending on the target's apparent size.\\
Because of the fast motion of the satellites and the limiting magnitude of the spectrographs, high-resolution instruments are essentially blind to most satellites. For lower resolution spectrographs, the effect on the measured astrophysical parameters depends strongly on the signal-to-noise of the exposure, longer exposures on brighter targets being the least affected at $\lesssim 1$\%. \\
Satellites that are brighter and/or higher than the constellation satellites, while less numerous, can also contaminate spectra.}
{Satellites, either from a large constellation or not, have the potential to contaminate spectra. While the fraction of affected spectra is likely to remain low even with a number of satellites about $100\times$ larger than today, some of these contaminated spectra will be difficult to identify, as it is already the case with existing satellites and asteroids. The best mitigation is to ensure that their brightness is fainter than $V=7$, that satellites' absolute magnitude $V_{\rm 1000 km}$ is also fainter than 7, and whenever possible to shoot multiple exposures.}


   \maketitle
%
%-------------------------------------------------------------------

\section{Introduction\label{sect:intro}}

While artificial satellites have been orbiting the Earth since 1957, their number has recently increased dramatically from around 1500 in 2015 to over 5000 today \citep{McD23a}. The main contribution to this increased is the result of large satellite constellations providing global internet coverage. Several constellation operators have plans to launch over 500\,000 satellites to low altitude orbits over the coming years \citep{McD23b}. The fate of many of these companies in uncertain. The maximum number of satellites that can be safely operated on low altitude orbit and the sustainability of such endeavour are a matter of debate \cite{law+22}. Nevertheless, it is clear that the number of low altitude satellites will continue to increase over the coming years. As of today, the main constellation operators are SpaceX, whose Starlink constellation already has over 4500 satellites in orbit, and OneWeb, with over 600 satellites launched.

This large ---and increasing--- number of satellites will have a series of environmental impacts \citep[][for a discussion]{law+22}, and will impact ground- and orbit-based astronomical observations in the optical and radio ranges. The International Astronomical Union has set up a ``Centre for the protection of the dark and quiet sky from satellite constellation interference'' (CPS\footnote{\url{https://cps.iau.org}}) to bring the astronomical communities, satellite operators, policy makers and community together, and consolidate and share data and information related to these interferences. A series of international conferences ---SatCon-1, \cite{SatCon1}, and -2, \cite{SatCon2}, Dark and Quiet Skies 1, \citet{DQS1}, and 2, \cite{DQS2}--- took place over the past years to exchange information, raise awareness, discuss policies across all stakeholders; their reports constitute very good reviews of the issues at hand. These reports also discuss the impacts on the environment and on radio astronomy, as well as mitigation methods.

In previous papers \citep{HW20, BHG22}, we evaluated and quantified the impact of satellite constellations on astronomical observations in the optical, using simulated plausible future constellations and a series of representative instrumental set-up. While many parameters must be considered in the simulations (number of satellites and their altitude, location of the observatory, time of the observations, type of instrument...), a simplistic 1-line summary of these simulations could be ``on average, few percent of the observations will be lost during the first and last hours of the night''. Measurements of the impact on actual observations are being evaluated and published, for instance \cite{mro+22} characterised the impact of the existing constellations on the Zwicky Transient Facilty (ZTF), a wide-field imaging telescope. They observe an increase in the fraction of contaminated images as the number of satellite increases, and infer than when the constellations will have grown to about 10\,000 satellites, ``essentially all ZTF images taken during twilight may be affected'', reaching about 4 satellite streak per image when the entire Starlink constellation will be completed (42\,000~satellites), nevertheless resulting in a data-loss of a fraction of a percent.

In imaging {observations}, a satellite leaves a tell-tale tail. While the apparent brightness of that tail can vary widely depending on the intrinsic magnitude of the satellite, on its apparent angular velocity, on the exposure time, pixel size, seeing, size and sensitivity of the telescope, the signature of the satellite is clearly visible down to extremely low resulting signal-to-noise (S/N) ratio. For spectroscopic {observations}, on the other hand, the situation is completely different. 
\begin{itemize}
    \item Integral Field Units (IFU) are spectrographs that combine imaging and spectroscopy to produce spatially resolved spectra of a two-dimensional region of the sky. The resulting data cube (2 spatial and one spectral dimensions) can be combined to reproduce an image of the sky, where a satellite trail will be apparent. For IFUs, the critical factor will therefore be sensitivity of the system.
    \item Slit spectrographs isolate a small, quasi uni-dimensional region of the sky. The width of the slit is typically matching the angular resolution of the telescope, of the order of $1''$ for classical telescopes, down to a few tens of milliarcseconds for telescopes equipped with adaptive optics. The slit can be ``long'' that is, of the order of an arcminute, typically for low-resolution spectrographs, or ``short'', a few arcseconds, usually for high-resolution, multi-order echelle spectrographs. A satellite crossing the slit will, if bright enough, cause a secondary spectrum. If it crosses close to or on top of the main science target, a careful examination of its spatial profile might reveal the contamination. 
    \item Fibre spectrographs collect the light in a tiny aperture, with a diameter of a few arcseconds at most, and feed it to a spectrograph via an optical fibre. In some instruments, a large number (thousands) of fibres can be positioned in the focal plane to observe many targets simultaneously. \B{While the absolute position of each fibre on the sky is well known, the data from the fibre have absolutely no resolved spatial information. Therefore, a satellite crossing a fibre can be identified only spectroscopically from the data themselves, or with some external assistance that flags whether a fibre exposure was contaminated.}
\end{itemize}

\section{Magnitudes\label{sect:mag}}
\subsection{Observed magnitudes}\label{sect:obsmag}
%----------------
\begin{table}
      \caption[]{Magnitude of constellation satellites}
         \label{tab:mag}
    \begin{tabular}{lccrrl}
    \hline
    Constellation & Band & $M_{\rm 1000km}$& $a$ & $M_a$ & \\
                 &      &                 & {[km]} &       & \\
    \hline\hline
    Starlink Gen. 1 
          & Vis& $	6.1	\pm	0.04$    & 550	&	4.8	&	1\\
    ~~Original (2019)   
          & $g'$&$ 6.46\pm 0.02$    & 550  &   5.16 & 2 \\
          & $r$& $ 5.9 \pm 0.02$&       550 &   4.6 & 2 \\
          & Vis& $ 5.9 \pm	0.45	$&	550	&	4.6	&	3\\
	       &$g'$& $ 6.4	\pm	0.5	$&	550	&	5.1	&	4\\
	       &Vis&  $	5.9	\pm	0.02	$&	550	&	4.6	&	5\\
          &$g'$& $ 6.7 \pm 0.05 $ & 550 & 5.4 & 6 \\
    \\
    Starlink Gen. 1 
       & $g'$ & $6.2\pm 0.04$ & 550 & 7.5 & 6\\
    ~~DarkSat 
       & Vis  & $6.4\pm0.2$ & 550& 5.1 & 7\\
    ~~(2019) \\
    \\
    Starlink Gen. 1 
          &Vis & $ 7.0 \pm 0.04 $ & 550 &	5.7	& 1\\
    ~~Visor (2020--21) 
	       &$g'$& $ 8.1 \pm 0.03 $ & 550 & 6.8 & 2 \\
          &$r$ & $ 7.6 \pm 0.03 $ & 550 & 6.3 & 2\\
          &Vis & $ 7.2 \pm 0.8  $ & 550 &	5.9	& 3\\
	       &$g'$& $	7.0	\pm	0.5	 $ & 550 & 5.7 & 4\\
	       &Vis & $	7.2	\pm	0.04 $ & 550 & 5.9 & 7\\
        \\
    Starlink Gen. 1 
         &Vis & $	6.34\pm	0.05$&	550	&	5.0	& 1\\
    ~~LaserSat, \\
    ~~post-visor \\
    ~~(2022)\\
    \\
    Starlink Gen. 2
         & Vis	&$	7.9	\pm	0.09	$&	530	&	6.5	&	2\\
    ~~Mini (2023) \\
    \\
    OneWeb	(current) &Vis & $	7.2	\pm	0.03$ &	1200	&	7.6	& 8	\\
    \hline
    \end{tabular}

{\footnotesize 
{\bf Notes:} The year in the Constellation column refers to the period during which that model was launched. $M_{\rm 1000km}$ refers to the apparent magnitude normalised to a distance of 1000~km between the observer and the satellite; for some references, the dispersion is estimated over normalised magnitudes while for others it is estimated over measured magnitudes; $a$ is the typical altitude of the satellites in the constellation, when in operation; $M_a$ is the apparent magnitude of a satellite at zenith, derived from $M_{\rm 1000km}$. \\
{\bf References: }
1: \citet{mal+23a}, 3: \citet{mro+22},
3: \citet{mal21b}, 4: \citet{bol+21}, 5: \citet{mal20b}, 
6: \citet{tr+20}, 7: \citet{mal21c}, 8: \citet{mal20a}.
}
\end{table}
%----------------

A satelllite must be illuminated by the Sun to be visible at optical wavelengths. As night falls, the Earth's shadow covers more satellites, starting with those in the East and at lower altitudes. See \citet{McD20} or \citet{BHG22} for more details. Radio observations, which are beyond the scope of this paper, are still affected by satellites in the Earth's shadow, as the satellites are radio emitters. Similarly, they emit thermal radiation detectable by mid-infrared instruments \citep[see][]{BHG22}. Satellites can also cause short eclipses by passing in front of the observed star; that effect was found to be extremely small to negligible \citep{HW20}. 

A satellite's magnitude depends on several factors: its size or cross section, its albedo (or more generally its bidirectional reflectance distribution, BRDF), the angles between the sun, the satellite, and the observer (including solar phase effects, and possible specular flares), the distance to the observer, and the elevation above the horizon (accounting for atmospheric extinction). Table~\ref{tab:mag} and references therein summarise the characteristic magnitudes of the main constellations, obtained by dedicated observers. The measurements also help to build brightness models of different spacecrafts \cite[e.g.][]{mal+22, kra+22,fan+23}. 

It is interesting to note that the efforts by SpaceX to make their satellites fainter is paying: their ``VisorSat'' design reduced their brightness by $\sim1$\,mag, and the change from their Generation 1 to Generation 2 made the satellites even fainter in spite of a much larger cross-section \citep{mal+23b}. This was achieved combining a novel specular coating causing the sunlight to be reflected to space (so it does not reach any observer on Earth) and an operation scheme that hides the illuminated solar panels behind the bus of the satellite. These improvements follow the SatCon workshop's recommendations to keep the satellites apparent magnitude $V$ and magnitude normalized to 1000km $V_{\rm 1000km}$ both fainter than magnitude 7 to make them invisible to the naked eye and not saturate the Vera Rubin Observatory camera \citep{tys+20}.

\subsection{Effective magnitudes}\label{sect:effmag}
%----------------------------------------------------------------- 
   \begin{figure}
   \centering
   \includegraphics[width=8cm]{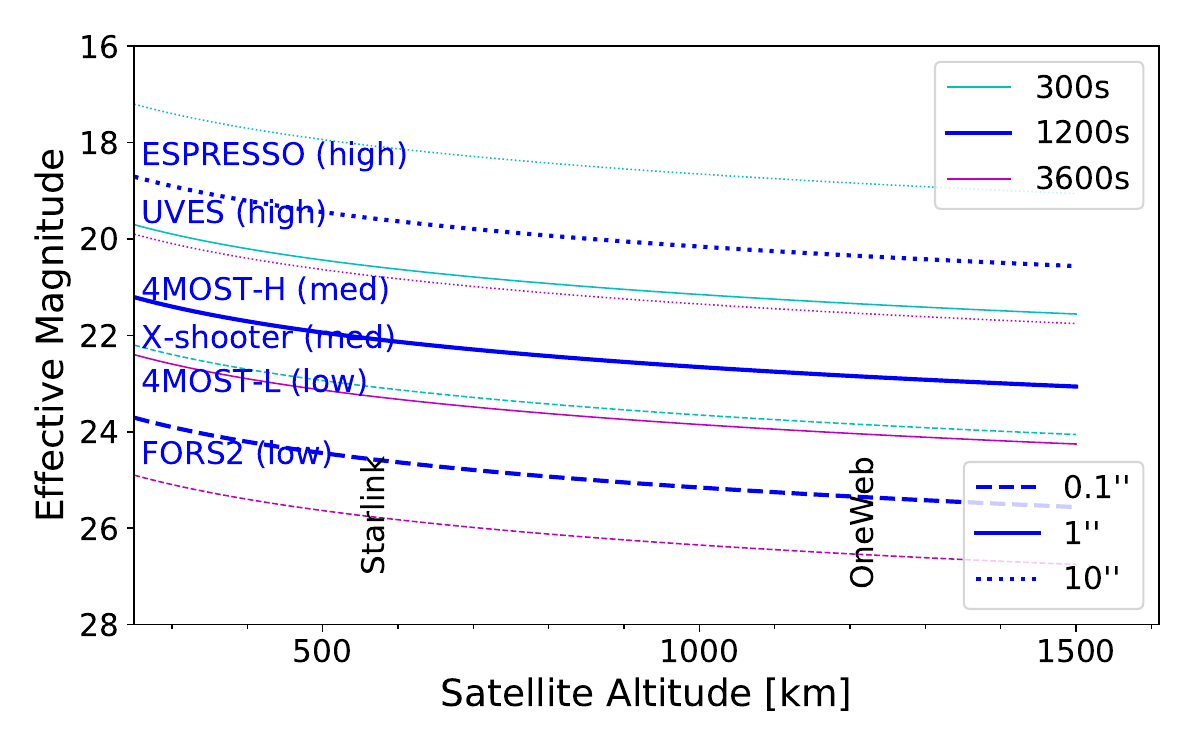}
      \caption{Effective magnitude $M_{\rm eff}$ of a satellite with $M_{\rm 1000km}=7$ as a function of its altitude, for various exposure times (colours) and spectrograph aperture width (line style). $M_{\rm eff}$ does not vary much with the elevation of the satellite above the horizon. Typical altitudes of the Starlink and OneWeb satellites are indicated. 
      The detection limit of a series of ESO spectrographs (for a 1200s exposure) is indicated; an instrument will only be sensitive to satellites brighter than that level.
      }
         \label{fig:effmag}
   \end{figure}
%-----------------------------------------------------------------

Because of its fast apparent velocity $\omega_{\rm sat}$, the satellite will cross the spectrograph's spatial resolution element (the fibre,  slit width, or IFU pixel) in $t_{\rm eff} = r / \omega_{\rm sat}$, where $r$ is the angular size of the resolution element on sky, of the order of 1$''$.  The value of $\omega_{\rm sat}$ is the angular orbital motion (related to the satellite altitude by Kepler's law) projected on the observer's sky. The derivations are given in \citet{BHG22}, Appendix~A. Typical apparent angular velocities are of the order of 10 to 60$\degree$/min, higher velocities being observed at high elevations. This results in $t_{\rm eff}$ of the order of 1~ms.

No matter how long the actual exposure time $t$ is, the satellite will only be exposed during $t_{\rm eff}$. Consequently, the contamination from the satellite will be more diluted over longer exposure times. It is useful to introduce the concept of {\em effective magnitude}, $M_{\rm eff}$, the magnitude of a static point source that would create a signal at the same level as that of the satellite, during the whole exposure time: 
\begin{equation}
    M_{\rm eff} = M_{\rm sat} - 2.5 \log( t_{\rm eff} / t ), \label{eq:effmag1}
\end{equation}
or
\begin{equation}
    M_{\rm eff} = M_{\rm sat} - 2.5 \log( r / \omega_{\rm sat} t ). \label{eq:effmag2}
\end{equation}
Interestingly,  $t_{\rm eff}$ and $M_{\rm sat}$ are both varying with the elevation of the satellite. They compensate each-other for satellites moving only in elevation, resulting in $M_{\rm eff}$ being fairly constant with the satellite elevation (see \cite{HW20}, Fig.10).

$M_{\rm eff}$ is useful to quickly determine whether a given spectrograph will be affected or not, comparing it to the limiting magnitude of the instrument for the considered configuration and exposure time. Figure~\ref{fig:effmag} plots $M_{\rm eff}$ for various combinations of exposure times and aperture sizes.

\section{Contamination probability\label{sect:proba}}

%----------------------------------------------------------------- 
   \begin{figure*}
   \centering
   \includegraphics[width=5.7cm]{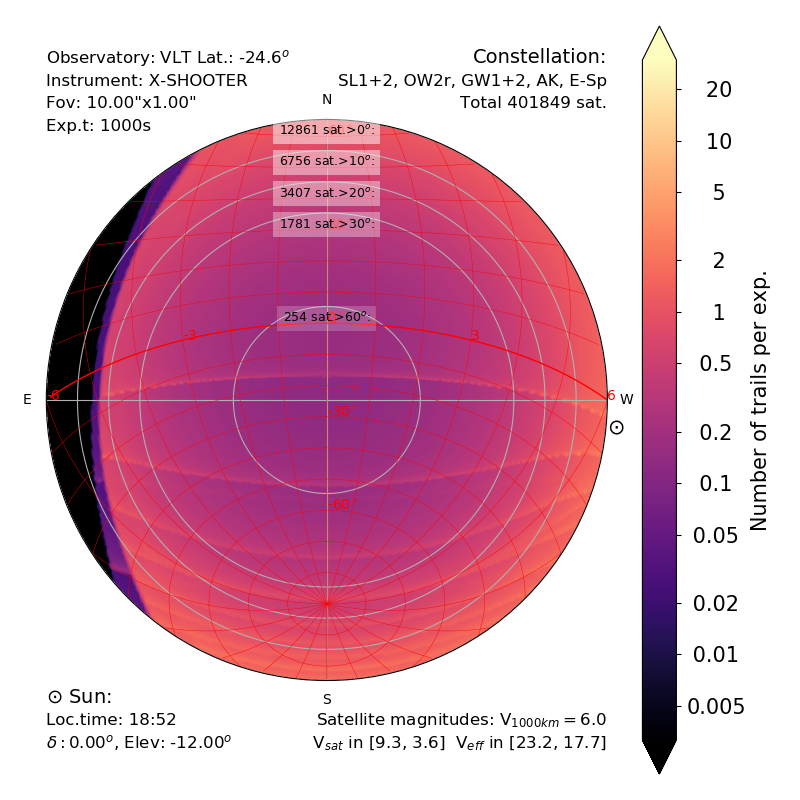}
   \includegraphics[width=5.7cm]{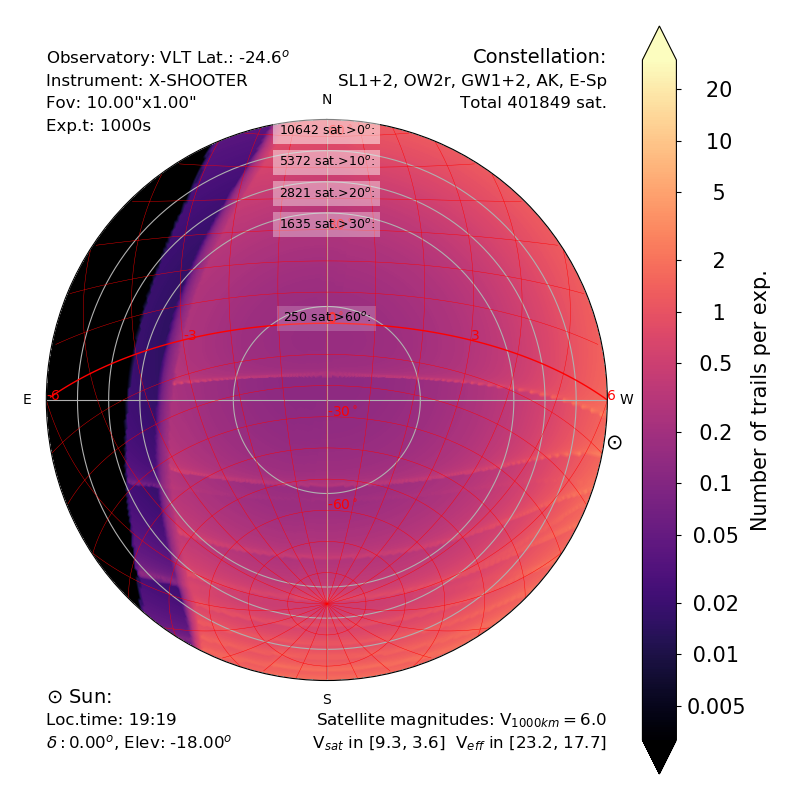}
   \includegraphics[width=5.7cm]{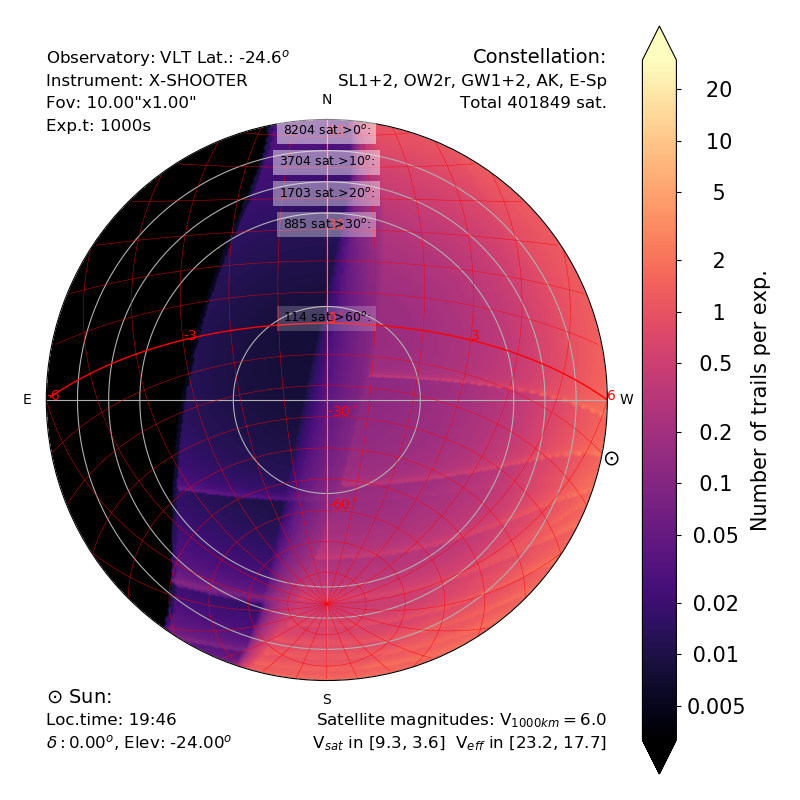}
   \includegraphics[width=5.7cm]{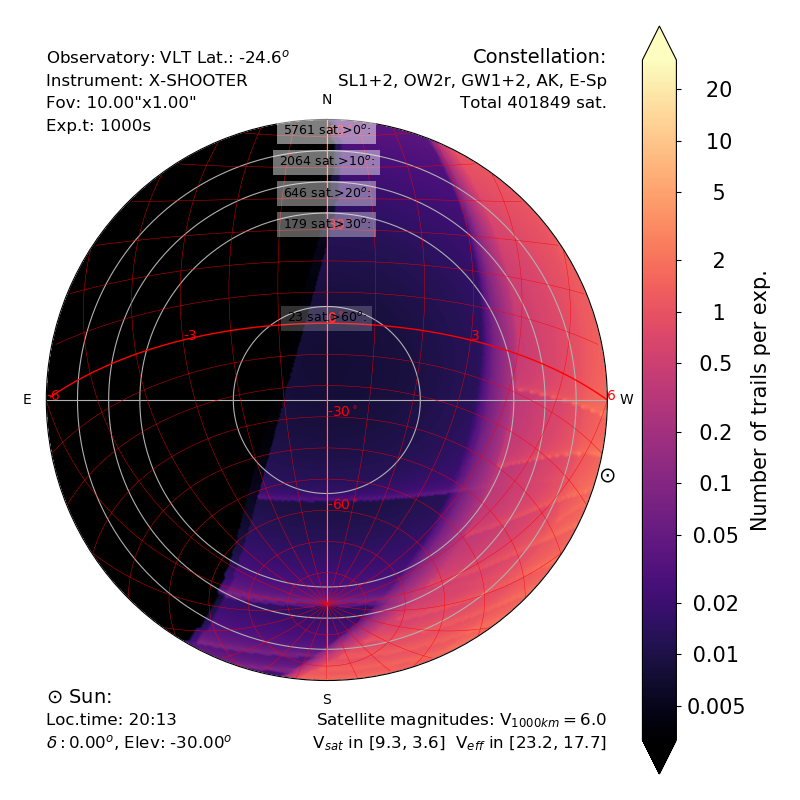}
   \includegraphics[width=5.7cm]{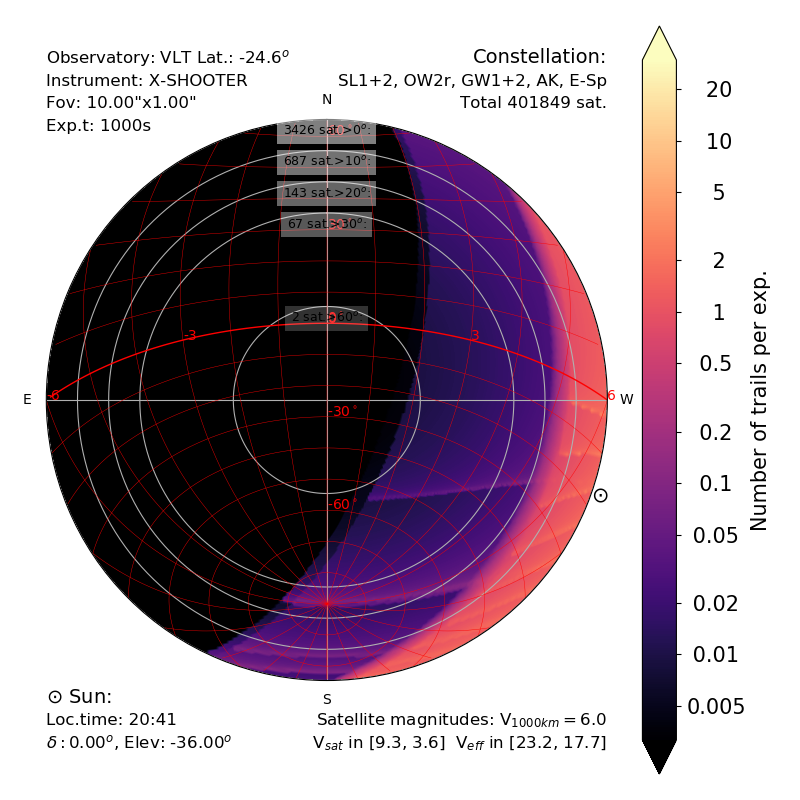}
   \includegraphics[width=5.7cm]{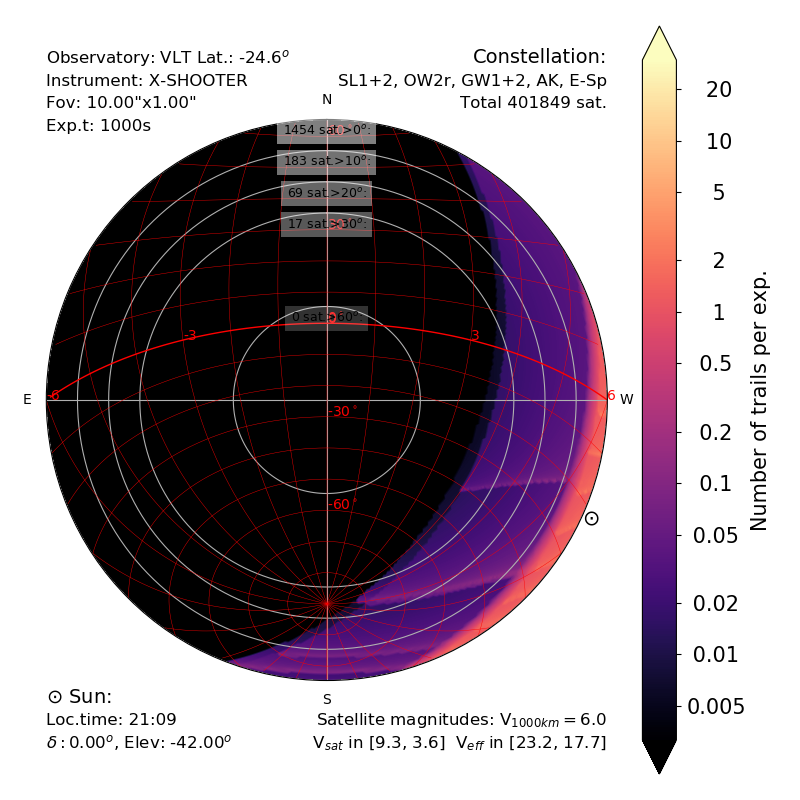}
   \includegraphics[width=5.7cm]{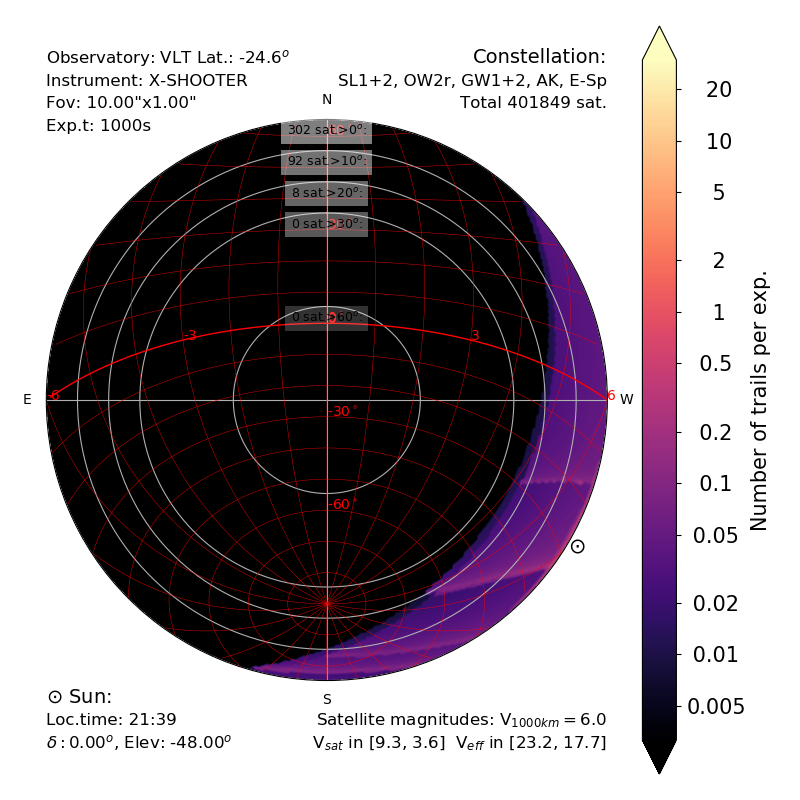}
   \includegraphics[width=5.7cm]{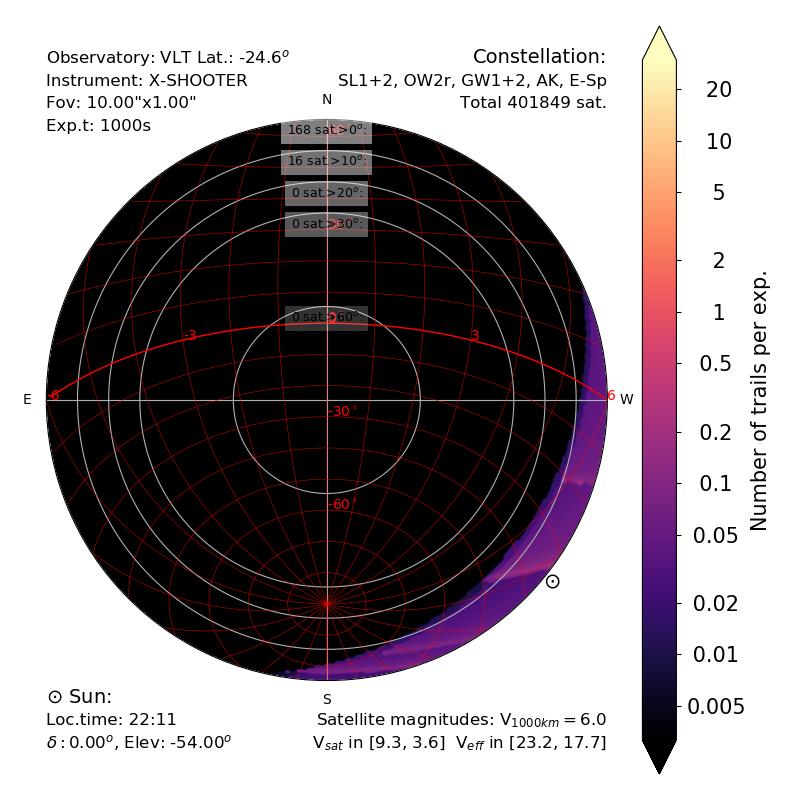}
   \includegraphics[width=5.7cm]{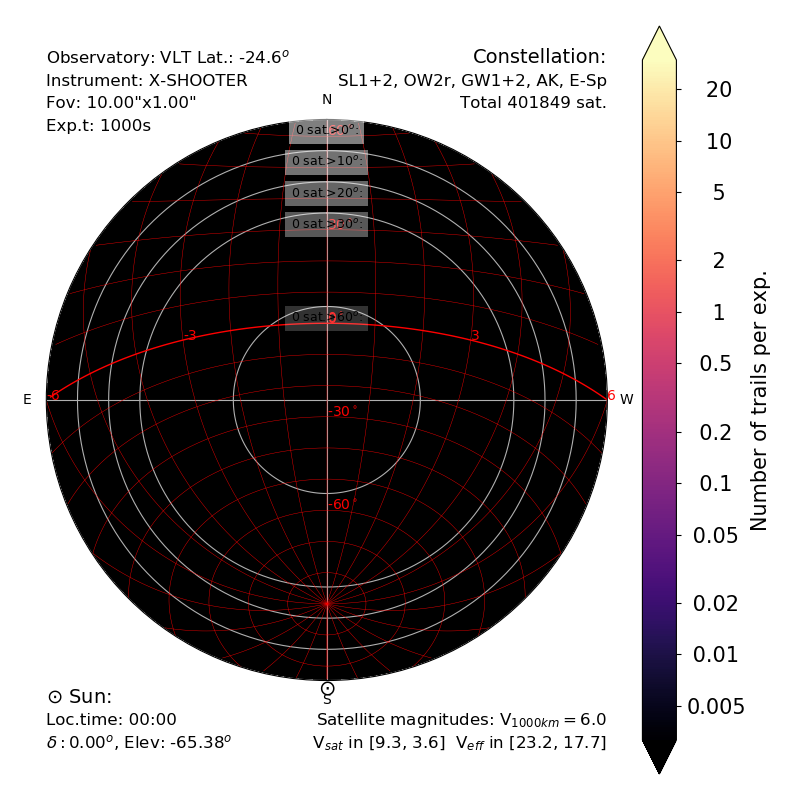}
      \caption{Maps of the sky above Paranal from twilight until mid-night; 
      the colour scale indicates the number of satellite trails crossing an X-Shooter 10$''$-long slit during a 1000\,s in that direction in the sky; \B{the text in the figure indicates that the simulations were performed for the VLT (at latitude $-24.6\degree$) for X-SHOOTER with a slit of $10\times 1''$ and an exposure time of 1000\,s (top left text); the constellations used for the simulation include 401\,849 satellites from Starlink, OneWeb, Guo Wang, Amazon Kuiper, and E-Sat (top right). The bottom-left text gives the local time and the declination ($0\degree$) and elevation of the Sun (from $-12\degree$ to $-54\degree$ with a step of $6\degree$,  and for mid-night at a solar elevation of $-65\degree$; finally, the bottom-right text indicates that the absolute magnitude of the satellites is set to $V_{1000{\rm km}}=6$, and gives the conversion into visual (in the range [9.3, 3.6]) and effective magnitudes (in [23.2, 17.7] for this geometry).    }
      }\label{fig:skymap}    
    \end{figure*}
%-----------------------------------------------------------------
%----------------------------------------------------------------- 
   \begin{figure}
   \centering
   \includegraphics[width=8cm]{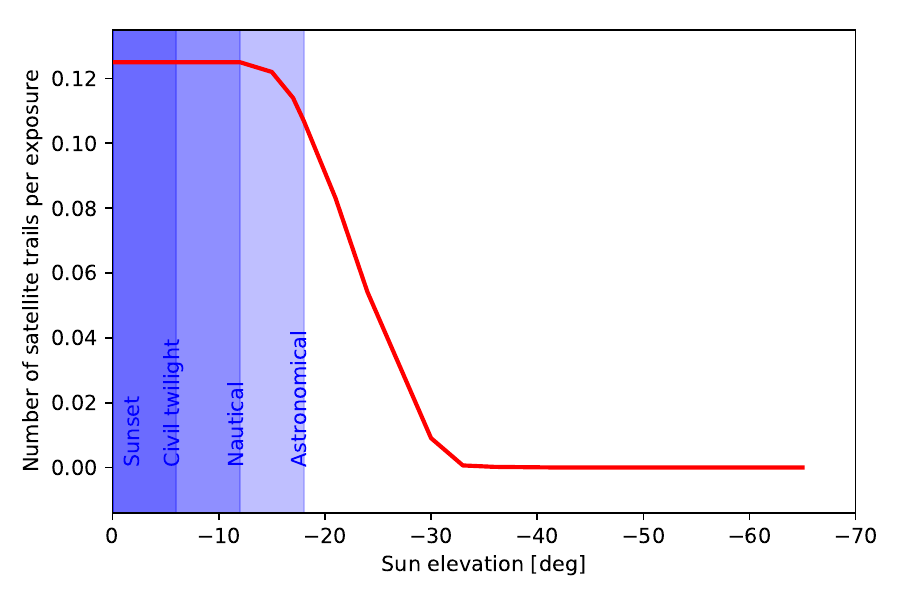}
      \caption{Number of satellite crossing the $10''$-long slit during a 1000s X-Shooter exposure, at zenith, as a function of the elevation of the Sun. The simulation considers over 400\,000 satellites. In first approximation, the number of satellite crossings scales with the number of satellite, exposure time and slit length.}
         \label{fig:losses}
   \end{figure}
%-----------------------------------------------------------------

The probability of a satellite contaminating an exposure is a function of the duration of the exposure, the size of the spectrograph aperture, and the density of illuminated satellites in the direction of observations. The latter can be estimated analytically from the parameters of the constellation (number of satellites, orbit inclination and altitude) using the formalism described in \citet{BHG22}. 
The source code is available on GitHub\footnote{\url{https://github.com/ohainaut/SatConAnalytic}}, and is deployed as a web-tool\footnote{\url{https://www.eso.org/~ohainaut/satellites/simulators.html}}

Using a worst-case scenario in which most foreseen constellations are fully deployed, totalling over 400\,000 satellites, we computed the number of satellite crossing a $10''$-long slit during a 1000~s exposure. The constellations used are those of SpaceX's Starlink (Generation 1 and 2), OneWeb (reduced), Amazon's Kuiper, Guo Wang, and E-Space (see \citet{McD23a} for details of the constellations). The actual details of the constellations are not important: globally, they constitute a representative population of what could be expected in the 2030s. The observations are modelled to take place from ESO's VLT in Chile (latitude 24.6$\degree$ South). The latitude does not change much the conclusions when reported to the elevation of the Sun. The simulations were run from twilight to midnight. Their results are displayed in Fig.~\ref{fig:skymap}.

The shadow of the Earth is clearly visible, engulfing the satellites as the Sun drops below the horizon, starting with the satellites on the lowest orbits. The horizontal (East-West) cusps in the distributions are caused by the edges of the constellation, where satellites ``pile up'' while reverting their motion southward to northward. When the Sun reaches $24\degree$ of elevation below the horizon, only the highest satellites remain illuminated at zenith, and by $30\degree$, the sky above $30\degree$ of elevation (corresponding to an airmass below 2, where most astronomical observations take place) is clear of illuminated satellites. The number of satellite trails for an observation at zenith is also represented in Fig.~\ref{fig:losses}. At the end of twilight, one can expect 0.1 trail per 1000~s exposure (or, in other words, one exposure in ten will be contaminated). That number drops to 0.05 (one in twenty) when the Sun reaches $24\degree$ below the horizon, and is $\sim$0 when the Sun is below 30$\degree$. Overall, the average over the astronomical night is 0.02~trail per 1000\,s exposure (one in 50).

These values scale with the number of satellite (here, 400\,000, i.e. about 100 times today's population), with the length of the slit ($10''$) and the exposure time (1000\,s). As an example, today's probability that a $3''$ fibre be hit by a satellite trail is $\sim6\times 10^{-5}$. %0.02 * 3/10 / 100

A satellite crossing the $10''$ slit does not automatically render the observation useless: depending on the object and on its spatial extension, it is plausible that the satellite will ``miss'' the target, contaminating the sky next to the object. Conservatively estimating that the width of the contamination is $5''$ (satellites are resolved by most professional telescopes, see \citealt{tys+20}), half the satellites would ``miss'' a point source in the $10''$ slit.

The case of giant multiplexed fibre spectrograph is worth discussing: while the number of satellites crossing the large field of view of the instrument can be very large (several tens), the focal plane is only sparsely populated by fibres. \citet{kov+23} have estimated that for the LAMOST spectrograph ($5\degree$ field of view populated with 4000 fibres), each satellite would hit on average 0.43 fibre. A similar estimation for the 4MOST spectrograph ($2.3\degree$, 2436 fibres), each satellite would contaminate 1.3 fibre on average.

The effect of the contamination will depend on the brightness of the satellite. This is discussed in next section.
\section{Contamination effects\label{sect:effects}}

\subsection{Low/med and high resolution}

Spectrographs can be broadly divided according to their spectral resolution, from low (with $R = \delta\lambda/\lambda$, where $\delta\lambda$ is the wavelength extent of a resolution element, in the 100--1000 range), where the dispersion is performed by a prism or a grism, medium ($R$ in 1000-10\,000) and high ($R$ in $10^4 - 10^6$, using echelle grating typically with a cross-dispersor). At constant telescope diametre and exposure time, the amount of light reaching the instrument is the same, so that amount is dispersed over a larger number of pixels at increasing resolution, resulting in a lower number of photon per pixel and therefore in a higher photon noise. Additionally, the detector read-out noise (which is constant) may become significant. Finally, the higher complexity of high resolution spectrographs tend to make them less efficient than low resolution instruments. Overall, the limiting magnitude (defined, e.g., as the magnitude for which a $5\sigma$ S/N spectrum is obtained) of a spectrograph is brighter at higher resolution.

The detection limit for a series of ESO spectrographs are indicated on Figure~\ref{fig:effmag}. Their resolution class is given after the instrument's name. 
Combining these limits with the satellites' effective magnitudes, it is clear that high-resolution spectrographs will be blind to constellation satellites. Low/medium-resolution instruments, on the other hand, will be able to detect some to most of the constellation satellites.

\subsection{Simulations}
%----------------
\begin{table*}
      \caption[]{Spectra used in the simulations}
         \label{tab:spectemplates}
    \begin{tabular}{llrrcrrr}
    \hline
Type              &Designation          & RA         & Dec      & $V$ & Exp [s] &Epoch 
    & Program \\
    &&&&&&& Archive \\
\hline\hline
Solar Analogue    &Hip042885            & 08:44:24.12& -57:46:06.0 &   8.1& 25 & 2018-03-18
    & 60.A-9022(C)\\
     (used as satellite)&
     \multicolumn{7}{r}{ADP.2018-05-04T12:16:31.417}  \\
He-rich hot star  &HZ 1          & 04:50:13.62& +17:42:07.6 &  12.7& 300 & 2015-12-26 
     & 096.D-0055(A) \\
    \multicolumn{8}{r}{ADP.2017-05-12T10:19:01.556} \\ 
Planetary nebula  &G295.3-09.3 = WRAY 16-86         & 11:17:42.96& -70:49:28.8 & 14.0&   80 & 2022-05-27 
    & 108.23MQ.001 \\
    \multicolumn{8}{r}{ADP.2022-06-09T06:02:45.995}  \\
Quasar            &QSO J124957-015928       & 12:49:57.41& -01:59:29.8 & 18.2& 1680 & 2012-05-17
    & 189.A-0424(A)\\
    \multicolumn{8}{r}{ADP.2014-05-17T10:30:26.313} \\
\hline
    \end{tabular}
    
    {\footnotesize {\bf Notes:} 
    RA and Dec are the coordinates in ICRS, $V$ the $V$-band magnitude; Exp is the exposure time. \\The data can be retrieved using the Archive file reference; their provenance is indicated by their ESO program ID. }
\end{table*}   
%----------------------------------------------------------------- 
   \begin{figure}
   \centering
   \includegraphics[width=8cm]{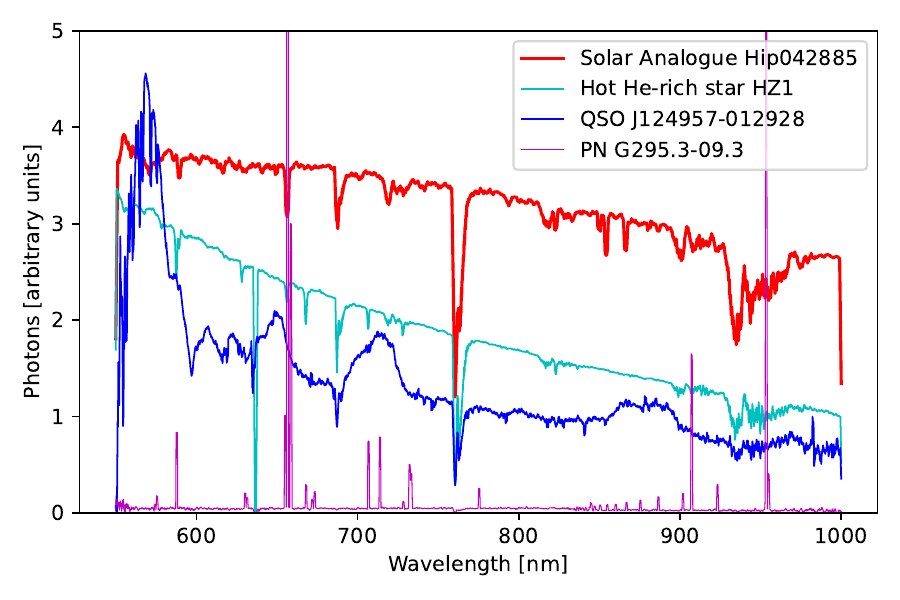}
      \caption{Spectra of a solar analogue used to represent satellite contamination and of three objects of interest. The spectra, in number of electrons, have been scaled arbitrarily and smoothed with a 10 pixel boxcar for clarity.}
         \label{fig:spectraex}
   \end{figure}
%-----------------------------------------------------------------
%----------------------------------------------------------------- 
   \begin{figure*}
   a.\includegraphics[width=8cm]{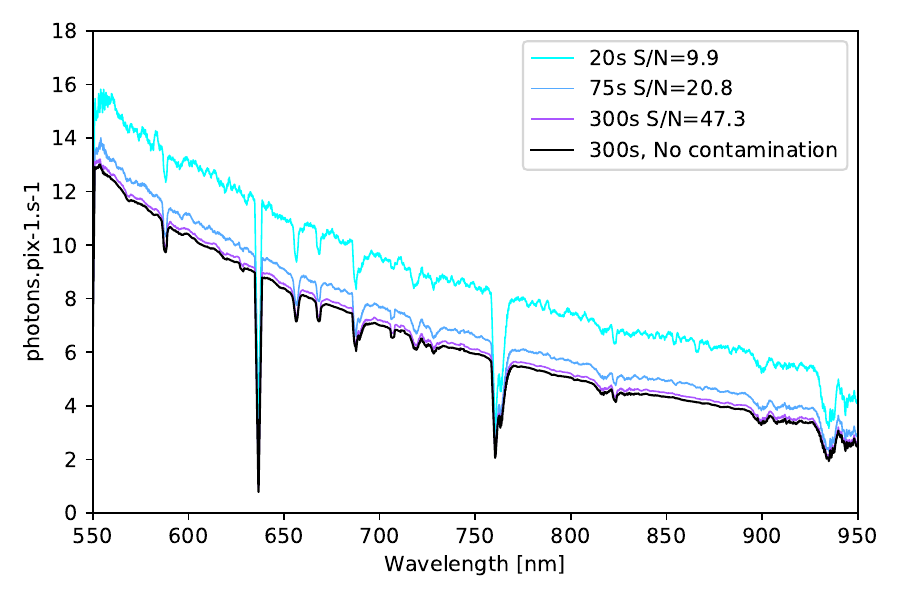}
   \includegraphics[width=8cm]{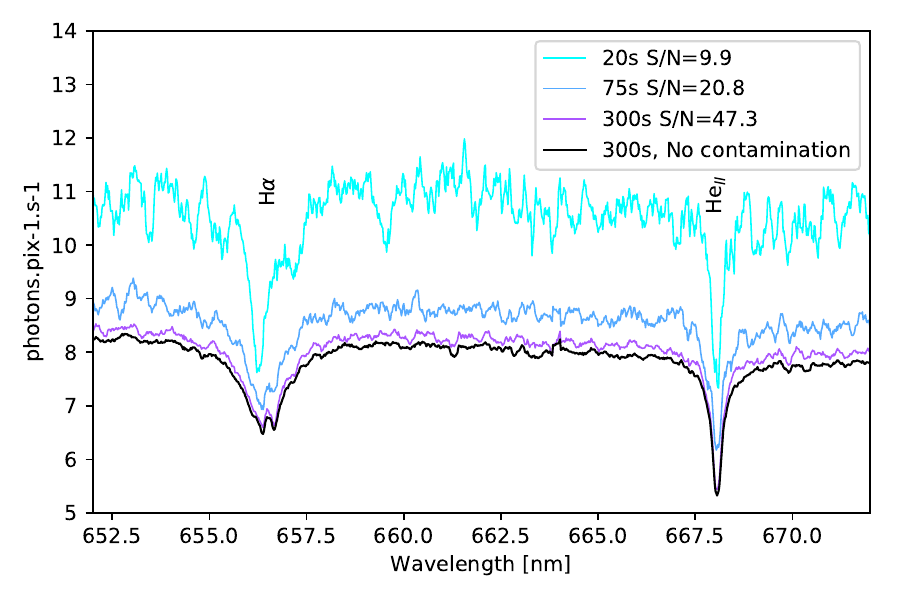}\\
   b.\includegraphics[width=8cm]{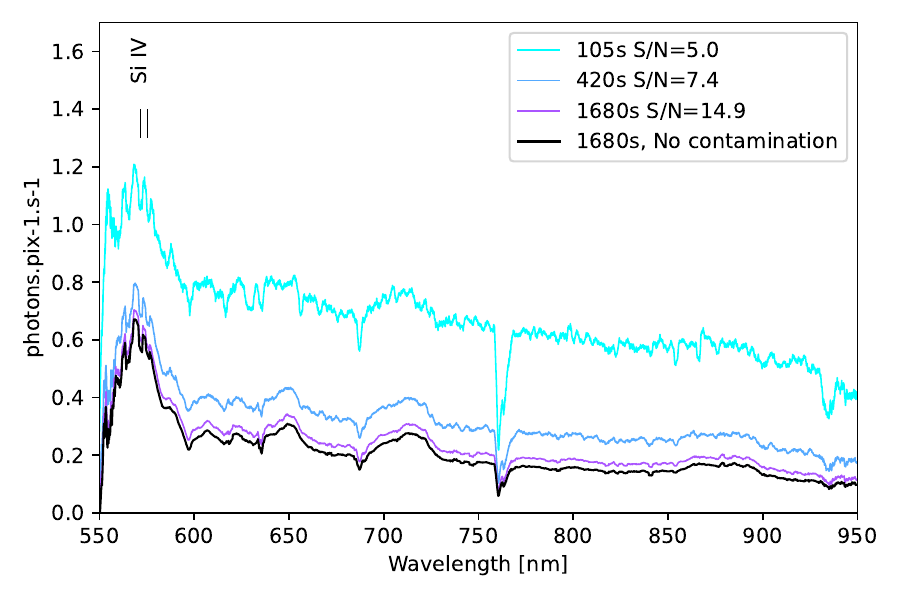}
   \includegraphics[width=8cm]{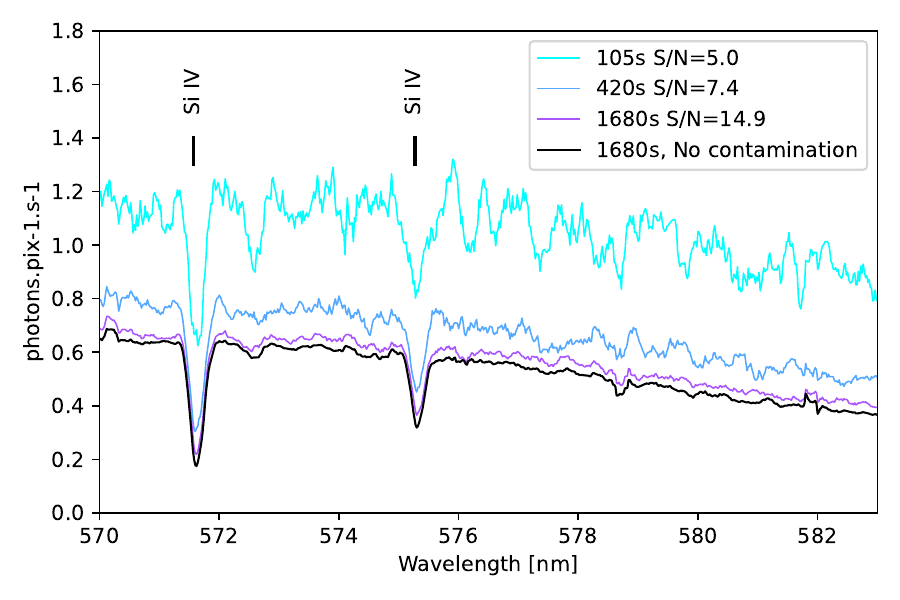}\\
   c.\includegraphics[width=8cm]{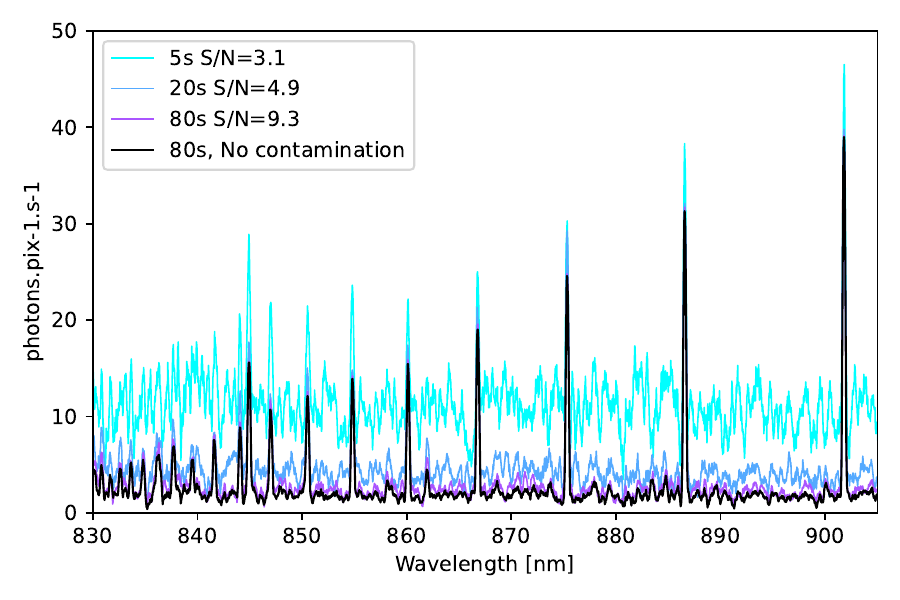}

      \caption{Spectra contaminated by a satellite trail (colour) compared to the original, uncontaminated spectra (black). a: hot star (full and detail); b: QSO (full and detail); c: planetary nebula. The full spectra were smoothed with a 10 pixel boxcar for clarity. 
         \label{fig:spectracontam}
         }
   \end{figure*}
%-----------------------------------------------------------------

In case a spectrum is contaminated by a satellite whose effective magnitude is much fainter than the limiting magnitude, the contamination will be extremely small, essentially increasing the noise in the spectrum by a negligible amount.

In case the spectrum is contaminated by a satellite whose effective magnitude is much brighter than the object being studied, the contamination will be obvious and the observation is essentially lost.

The most critical and difficult case is that of a satellite whose effective magnitude is close to that of the target. The contamination will not necessarily be obvious, and the measurement obtained from that spectra will be affected. In order to evaluate this case in more details, a series of X-Shooter spectra \citep{xshooter} were selected. X-Shooter is a medium-resolution echelle spectrograph equipped with three arms optimized in the near-UV and blue, in the visible, and in the near-infrared. Data from the visible arm, which were used hereafter, have a spectral resolution of around 8\,000 over 550 to 1020~nm for a slit width of $1''$. The $5\sigma$ limiting magnitude for a 1h exposure is $V_{\rm AB} = 20.8$, so constellation satellites are well in reach of X-Shooter.

The ESO Science Archive was queried for data processed by the X-Shooter pipeline \citep{xshopl}; the instrumental signature is removed, resulting in sky-subtracted, flux and wavelength calibrated, ``science-ready'' spectra, with an estimate of the corresponding noise.

To represent the satellite contamination, a series of well exposed solar analogue stars were investigated. A 25\,sec spectrum of Hip~042885 was selected for its high signal-to-noise ratio. The magnitude of the contaminating satellite was set to $V=6$  as representative magnitude of current and upcoming satellites (see Table~\ref{tab:mag}). This is hopefully conservative, as recent satellites are fainter. The time to cross the resolution element was set to 2~ms, corresponding to $25\degree$/min for a resolution element of $3''$. This is a conservative lower limit: most satellites will move faster, and the typical X-Shooter slit width is $1''$,  therefore resulting in a weaker contamination. 

A series of representative objects was selected from the X-Shooter archive; they are listed in Table~\ref{tab:spectemplates} and their spectra plotted in Fig.~\ref{fig:spectraex}. The flux, originally provided in erg~cm$^{-2}$~s$^{-1}$\AA$^{-1}$, were converted in photons per pixel during the exposure time. The corresponding photon noise was checked against the estimate provided by the pipeline (which includes the photon noise and all other sources of noises). The contribution of the sky to the noise was estimated using a {synthetic sky spectrum generated with ESO's SkyCalc application\footnote{\url{https://www.eso.org/sci/software/pipelines/skytools/skycalc}} \citep{Noll+2012,Jones+2013}.} 
It was found much smaller than the other effects, and therefore neglected. 

To explore the contamination effects on spectra of various S/N and exposure times, the original spectra have then been scaled to a series of exposure times, also scaling the noise to the new exposure level.  For each spectrum, the contamination by a satellite passing in front of the target was then added, and its contribution to the noise accounted for. The exposure times were selected starting from the original exposure time, and decreasing by factors of 4 (which would result in halving the S/N in the absence of a satellite) until the S/N becomes too low to be useful. The resulting spectra are displayed in Fig.~\ref{fig:spectracontam}.

%%%%%%%%%%%%%%%%%%%%%%%%%%%%%%%%%%%%%%%%%%%%%%%%%%%%%%%%%%%%%%%%%%%%%%%%%%%%%%%%%%%%%%%%%%%%%%%%5
\subsection{Analysis of the simulations}
%\R{In all what follows, we list values (Eq.w, fluxes...) without error bars. Use the S/N to guesstimate the error (S/N = 20, flux = 2.5 so error = 0.5}

Satellite trails can affect spectra in different ways: raising the continuum, changing the shape, position and/or relative depth of absorption lines. Below we evaluate these different aspect on different objects and some of their astrophysical parameters.

\subsubsection{Hot star}

\begin{table*}
     \caption[]{Fluxes and equivalent widths for the contaminated $V=12.7$ hot star}
         \label{tab:star}
\begin{tabular}{rrrccrccr}
\hline
Exp.& Sat.& S/N &         \multicolumn{3}{c}{He~I (667.8~nm)}     & \multicolumn{3}{c}{Proportions} \\
  t. & Eff.& 	  & Cont. & Line & Eq.w. & Cont. & Line & Eq.w.  \\
{[s]} & Mag.&	      & $\times 10^{-18}$  &  $\times 10^{-19}$  & [m\AA] &     &      & \\
\hline\hline
\multicolumn{9}{l}{Contaminated}\\
%  5 & 14.5 &  7.9 &   3.18 & -1.31 & 140.6 & 1.52 & 1.06 & 0.70 \\ 
 20 & 16.0 & 16.3 &   2.37 & -1.26 & 181.0 & 1.13 & 1.01 & 0.90 \\ 
 75 & 17.4 & 35.4 &   2.17 & -1.24 & 195.5 & 1.04 & 1.00 & 0.97 \\ 
300 & 18.9 & 78.3 &   2.13 & -1.24 & 199.9 & 1.01 & 1.00 & 0.99 \\ 
\multicolumn{9}{l}{Not contaminated}\\
300 & --    & 90.0 &   2.09 & -1.24 & 201.6 &   1  & 1    & 1    \\
\hline
\end{tabular}
                 
{\footnotesize {\bf Notes:} 
Exp.t is the exposure time in seconds; Sat.Eff.Mag. the satellite effective magnitude (for a contaminating satellite with $V=6$ moving at 25$\degree$/min); S/N is the measured signal-to-noise ratio; 
Continuum and Line are the fluxes (in erg~cm$^{-2}$~s$^{-1}$\AA$^{-1}$) measure for the {wavelength range} 666.047\,nm--669.453\,nm); proportion lists the ratio of the quantities relative to the non-contaminated spectrum.
}
\end{table*}

Abundance measurements are a critical input to stellar astrophysics and often rely on measured {\em equivalent widths}. To evaluate the satellite contamination effects, the He\,I line at 667.8~nm was measured in the range [666.047, 669.453~nm], extracting the continuum level, the flux in the absorption line, and the equivalent width. The values are reported in Table~\ref{tab:star}, and a close-up view of the spectra is available at Fig.~\ref{fig:spectracontam}~a (right).

As expected, the satellite contamination raises the continuum level. As the contamination is independent of the exposure time, that effect is stronger for shorter exposures, ranging from 1\% for the original exposure time up to $\sim50$\% for the shortest exposure. Also as expected, the line flux is not affected by the satellite, because the solar spectrum contains only a weak absorption line in that wavelength range; the increase for the shortest exposure is a reflection of the poor S/N of that spectrum. Combining both the continuum and line flux, the equivalent width is also affected, at the 1\% level for the original exposure time, up to 10\% for the low S/N spectrum. 

In summary, the contamination effect is small (1--3\%) on the properly exposed spectra {for wavelength regions without strong lines in the solar spectrum}, but increases with decreasing S/N. The obvious conclusion is that projects requiring high precision should work with high-S/N data, {which is advisable even without satellite contamination}.

{Reliable {\em line positions} are important for a number of astrophysical analyses, like redshifts of galaxies, binary studies, abundance determinations.} The position of the H$\alpha$ line (656.3~nm) {in the hot star spectrum} is shifted, even at fairly low levels of contamination, {because the solar spectrum contains an H$\alpha$ line}. At shorter exposure times, the {\em line shape} and position becomes dominated by that of the satellite. {line shapes are for instance important to determine atmospheric parameters of certain stellar types like white dwarfs, but also to identify the possible presence of companions.
}

\subsubsection{Quasar}

\begin{table}
     \caption[]{Equivalent widths for the contaminated $V=18.2$ quasar}
         \label{tab:qso}
\begin{tabular}{rrrcccc}
\hline
Exp.  & Sat.& S/N & \multicolumn{4}{c}{Eq. width.} \\
  t.  & Eff.& 	  &  \multicolumn{2}{c}{139.376~nm} & \multicolumn{2}{c}{140.277~nm} \\
 {[s]}  & Mag.&	  &  [m\AA]     & Prop. & [m\AA]   & Prop.\\
\hline\hline
\multicolumn{5}{l}{Contaminated}\\
105   &  17.81 &  4.97   & 106.0 & 0.510 &  77.4  & 0.599 \\
420   &  19.31 &  7.41	 & 166.1 & 0.800 & 106.7  & 0.825 \\
1680  &  20.82 &  14.89	 & 191.2 & 0.921 & 119.5  & 0.925 \\
\multicolumn{5}{l}{Not contaminated}\\
1680  &   --   &  20.66	 & 207.6 & 1     & 129.2  & 1 \\
\hline
\end{tabular}

{\footnotesize {\bf Notes:} 
Exp.t is the exposure time in seconds; Sat.Eff.Mag. the satellite effective magnitude (for a contaminating satellite with $V=6$ moving at 25$\degree$/min); S/N is the measured signal-to-noise ratio; next columns list the equivalent widths and proportion to the uncontaminated case.
}
\end{table}

Chemical abundance measurements also play a role in quasar astrophysics and, more generally, cosmology. Our example quasar spectrum {\citep[from][redshift = 3.10265]{sac+23}} was obtained to characterise stars in the early universe, through measurements of a series of lines. Two of them, Si\,IV at rest wavelength 139.376~nm and 140.277~nm {(corresponding to observed air wavelengths 571.574\,nm and 575.271\,nm)} were found to be suitable for automatic equivalent width measurements, and therefore were used to evaluate the satellite contamination impact.

Even for the high-S/N exposure with the original exposure time, the equivalent widths are affected at the 8\% level. At lower S/N, the difference raises above 40\%.

\subsubsection{Planetary nebula}
\begin{table}
     \caption[]{Flux ratio for the contaminated $V=14.0$ and 19.0 planetary nebulae}
         \label{tab:pn}
\begin{tabular}{rrrrrrr}
\hline
Exp.&
     Sat.& S/N&        
     \multicolumn{2}{c}{Electron}&
     \multicolumn{2}{c}{Electron}\\
t.  & Eff.&&  \multicolumn{2}{c}{Density }&
     \multicolumn{2}{c}{Temp.}\\
{[s]} & Mag.&&   (S\,II)&
                         Prop.&
                             (N\,II)&
				                           Prop.\\
\hline\hline
\multicolumn{6}{l}{Original magnitude}\\
\hline
\multicolumn{6}{l}{Contaminated}\\
5   &14.5  &    3.1  &      0.440&0.972 &   23.198 &  1.022\\
%10  &15.2  &    3.8  &      0.446&0.986 &   22.947 &  1.012\\
20  &16.0  &    4.9  &      0.450&0.993 &   22.824 &  1.005\\
80  &17.5  &    9.3  &      0.452&0.998 &   22.732 &  1.001\\
\multicolumn{6}{l}{Not contaminated}\\
80  & --    &    9.3  &      0.453&1      &   22.703 &  1\\
\hline
\multicolumn{6}{l}{5 mag. fainter}\\
\hline
\multicolumn{6}{l}{Contaminated}\\
%42  & 16.8 &    3.70  &      0.297&0.655 &   30.117 &  1.327\\
166 & 18.3 &    3.9  &      0.412&0.909 &   24.097 &  1.061\\
666 & 19.8 &    5.3  &      0.441&0.973 &   22.915 &  1.009\\
2666& 21.3 &    9.3  &      0.448&0.989 &   22.637 &  0.997\\
\multicolumn{6}{l}{Not contaminated}\\
2666& --    &    9.3  &      0.453&1      &   22.703 &  1\\
\hline
\end{tabular}

{\footnotesize {\bf Notes:} 
Exp.t is the exposure time in seconds; Sat.Eff.Mag. the satellite effective magnitude (for a contaminating satellite with $V=6$ moving at 25$\degree$/min); S/N is the measured signal-to-noise ratio; the S\,II and N\,II columns list the flux ratio, and Proportion is the ratio relative to the non-contaminated case.}
\end{table}

The spectrum of a planetary nebula has been scaled and contaminated as described above. The original spectrum was then scaled to represent an object 100~times fainter, and the exposure time increased to compensate and obtain the same S/N as for the original, brighter object. This second spectrum was then processed as the original, decreasing the exposure time by factors of 4 and contaminating with the satellite. With these two template objects, we can cover a much broader range of exposure times.

The electron temperature and electron density are important parameters for the characterisation of a planetary nebula {and can be measured from the flux ratios of certain lines}. \citet[Section 5]{OF06} give a series of diagnostic lines whose flux ratio can be used to derive these parameters: the electron density is related to flux ratio of S\,II lines $f$(671.6~nm)/$f$(673.1~nm), and the electron temperature to the N\,II lines as ( $f$(654.8~nm)+$f$(658.3~nm))/$f$(575.5~nm). These lines' fluxes have been measured in the contaminated and uncontaminated spectra, and are listed in Table~\ref{tab:pn}.

The flux ratio measured on the contaminated exposure that have the same exposure time as the bright original, non-contaminated frame differ by only 0.2\%, resulting in a negligible difference in electron temperature and density. 

In the worst case --the fainter nebula with the shortest exposure time (166s )-- the S\,II ratio is 9\% lower than in the non-contaminated spectrum. Using the relation from \citet{OF06}, this would change the electron density from $\approx2500 {\rm cm}^{-3}$ to $\approx1800 {\rm cm}^{-3}$, i.e. an error of almost 30\%. Similarly, a $\sim5$\% change on the N\,II ratio would bring the electron temperature from $\approx$19\,000~K to $\approx$17\,200~K, i.e. an error of $\approx10$\%.

\subsubsection{Simulation summary and discussion}
%----------------------------------------------------------------- 
   \begin{figure}
   \centering
   a\includegraphics[width=8cm]{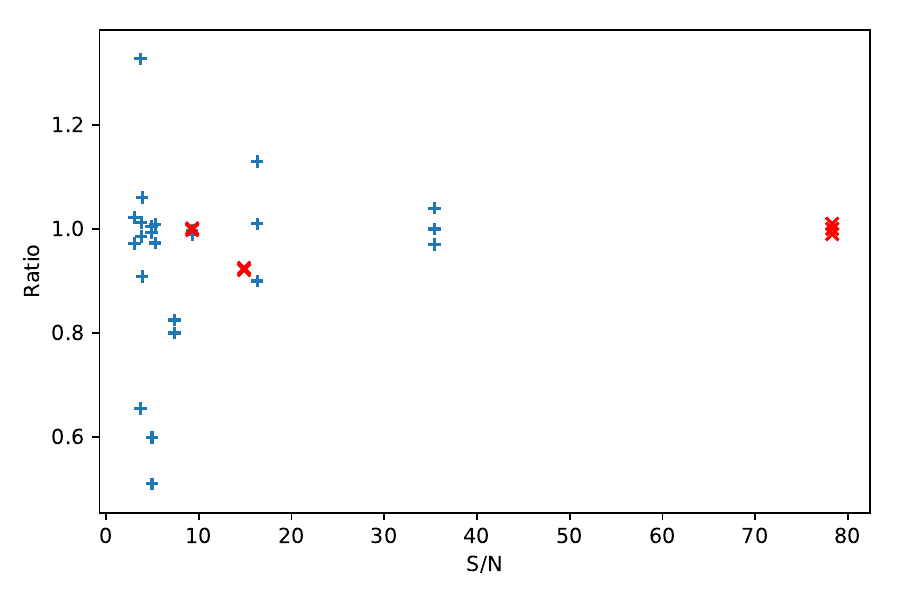}
   b\includegraphics[width=8cm]{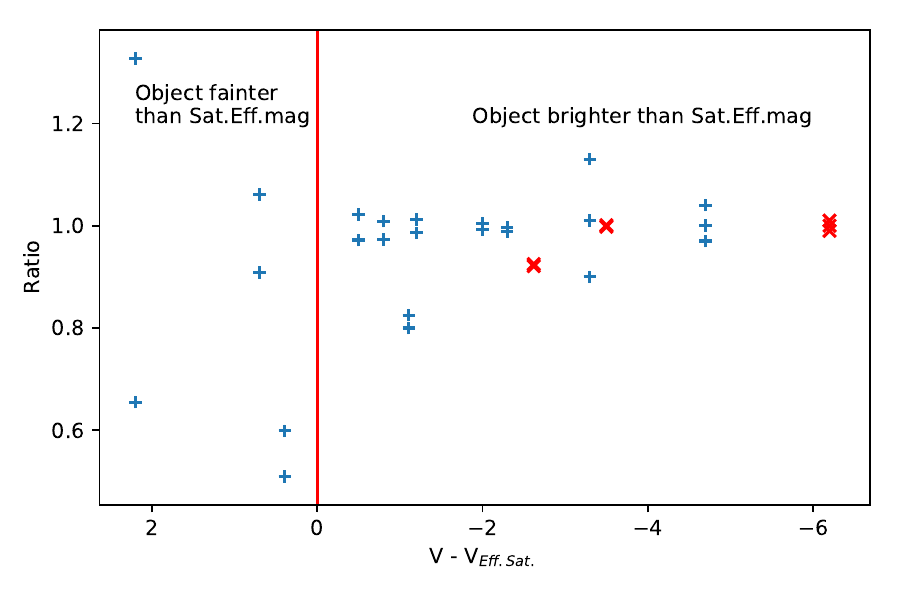}
      \caption{Ratio of contaminated measurement to uncontaminated, from Tables~\ref{tab:star}, \ref{tab:qso}, and \ref{tab:pn}, as a function of the spectrum S/N (a) and of the difference between the object's magnitude and the satellite's effective magnitude for the exposure. Red crosses mark measurements performed on exposures with the original exposure times, blue plus those where the exposure time was shortened.
      }
         \label{fig:ratios}
   \end{figure}
%-------------------
\B{The simulated satellite contamination spectra are presented for their illustrative value. The numerical values and fraction of contamination should be taken as a starting point to evaluate whether a given science case on a given instrument is at risk, and whether further investigation is needed.}

Overall, for spectrographs that are sensitive enough to detect constellation satellites, a rather obvious conclusion is that longer exposures with higher S/N on objects brighter than the effective magnitude of the satellite (or any combination of these condition) are less affected than shorter exposures on objects fainter than the {effective magnitude of the} satellite, with lower S/N. In other words, good spectra are less affected by satellites than worse spectra... This is illustrated by the plots in Fig.~\ref{fig:ratios}. With the pessimistic satellite characteristics used in these simulations ($V$=6, while most constellation satellites are expected to by  fainter, and angular velocity of 25$\degree$/s, while most constellation satellites are faster), measurements on exposures with high S/N are affected at the $\sim1$\% level. Overall, the contamination effects by constellation satellites are therefore expected to be small to negligible for all but the most stringent requirements. 

As science in general and interesting objects in particular tend to appear first in low S/N data, it is interesting to note that the S/N of the exposures four times shorter than the next one is higher than the expected 2 times smaller: the fixed contribution of the satellite becomes more and more dominant for shorter exposures. Therefore, a possible tell-tale sign of a contamination is that the S/N of the data is higher than expected for the magnitude of the object. In case of a weird or unexpected measurement, comparing the magnitude of the object to the effective magnitude of possible contaminators (using Fig.~\ref{fig:effmag}, or Eq.~\ref{eq:effmag1} or ~\ref{eq:effmag2}) may help decide whether a contamination is plausible. {To notice anything unusual in a measurement, however, a user needs to have some knowledge about the kind of observations (and instrument) and the expectation for that type of object, which may not be the case for Virtual Observatory users retrieving processed `science ready' data from a diverse selection of instruments.}

Quantitatively, combining the probability of contamination from Sect.\ref{sect:proba} (e.g. $\sim0.05$ for an 1000\,s X-Shooter twilight exposure with 400\,000 constellation satellites) with the effect of such contamination from this section (which considers satellites brighter and slower than the immense majority of constellation satellites), the impact constellation satellites on spectroscopic observations will be statistically small. Recognising the few contaminated spectra will require care.

The simulations considered satellites with $V=6$, which is on the bright end of constellation satellites. There are, however, much brighter satellites: the International Space Station is routinely observed around magnitude 0, the Hubble Space Telescope at $V\sim2$. Fortunately, these individual satellites are not a significant threat for spectroscopic exposures. Indeed, while the ISS crossing the slit during an exposure would ruin it, the probability for this to happen (during a 1000\,s exposure at zenith with a 1$''$ slit, as in Fig.~\ref{fig:losses}) is of the order of $10^{-7}$: one in several million spectra is expected to be contaminated by the ISS.
Recently, {\em AST SpaceMobile}\footnote{\url{https://ast-science.com/spacemobile-network/bluewalker-3/}} has announced their intention to launch a constellation of 243 very large satellites ($\sim64~$m$^2$) to provide direct cell phone connectivity. The first prototype, BlueWalker~3, was launched in Sep.~2022. The unfolded satellite was observed \citep{mal+23b} between magnitude 1.0 and 4.0 for 83\% of the measurements performed while the satellite was in operational attitude, corresponding to a mean $M_{\rm 1000km}= 3.1$. With such brightness, a BlueWalker~3 would ruin most low-resolution spectra and strongly contaminate medium-resolution ones. If 243~satellites are indeed deployed, 3--10 of them would be easily visible to the naked eye during twilights, but fortunately the probability for one of them to ruin an exposure would still be very low ($\sim10^{-5}$ using the same conditions as above).

Starlink and OneWeb satellites are launched by groups of 20--60 satellites. Once released from the launcher, they form a spectacular ``train'' of satellites that slowly disperses as the satellites drift apart, until their orbits are adjusted to their final configuration. During that period, the satellites are typically on lower orbits and their attitude is not the operational one, resulting in brighter magnitude. The early Starlink trains were particularly bright; this has been improved as SpaceX has changed their procedures. These bright trains caused impressive ``photobombing'' that raised the awareness of the community about the issue of satellite pollution. Considering 400\,000 satellites with an operational life-time of 5~years, replenishing the constellations would require 3--4 launch/day (with 60 satellites each). If it takes about one month for these satellites to reach their operational altitude and attitude, we would have $\sim6000$~objects on these low transfer orbits. Scaling their probability of interaction with a spectrograph from Fig.~\ref{fig:skymap} and \ref{fig:losses} results in one contaminated X-Shooter spectrum every few thousand exposure. Notwithstanding all other considerations about sustainability and orbital crowding, it is therefore critical that the satellite operators strive to keep the orbital transfer phase as short as possible, and to keep the satellites as faint as possible also during that phase, within the general guideline to keep them fainter than $V=7$.

While a satellite on an orbit higher than the constellations is typically fainter, its apparent velocity slower, and both effects partially cancel each-other, resulting in the effective magnitude becoming slowly fainter with increasing altitude (see Fig.~\ref{fig:effmag}). Therefore, the few tens of thousand of satellites present on higher orbits cannot be neglected. This was painfully illustrated by the case of the gamma-ray burst at redshift 11 candidate that \cite{mic+21} identified as the contamination by a Proton rocket debris. The satellite, on a very elliptical orbit, was at 15\,000\,km from the observatory, and caused a $M_{\rm eff}=19.2$ contamination on the 179\,s exposure through a $0.9''$ slit. \cite{mic+21} report that the apparent angular velocity of the object was $\omega = 36''$/s. Using these values and a derivation similar to that of Eq.~\ref{eq:effmag1}, they obtained $M=9.7$, which converts in $M_{\rm 1000km} \sim3.7$. This illustration justifies the recommendation that satellites should have $V>7$ {\em and} $V_{\rm 1000km}>7$.

Even higher up, asteroids (and comets) are known to cross exposures at inopportune times and position, causing (stellar and extra-galactic) astronomers to refer to them as ``vermin of the sky.'' With apparent velocities in the range of $\sim1''$/h for trans-neptunian object to $1''$/s for near-earth asteroids, and with absolute magnitudes $M_{\rm 1au}$ ranging from 3 to 27 (the use of $M_{\rm 1000km}$ is not  practical beyond earth's orbit), asteroids can be relied upon to ruin an exposure, as illustrated by \citet{haa+24} or \citet{rod+19}. Furthermore, only a fraction of these objects are known and catalogued. \citet{ste+21} discuss the cases of 27 single-exposure transient from over 81\,000 MOSFIRE spectra, concluding they were likely caused by solar system objects or high-altitude satellites. 

To complete the list of possible contamination sources, one must add meteor, air planes, cosmic rays, a series of instrumental features that can mimic a detection, and even pipe smokers \citep{AM67}. Overall, satellites are joining the list of troublemakers.

\section{Mitigation}\label{sect:mitig}
In \citet{BHG22}, we listed a series of possible mitigation methods. In summary, 
at the level of the satellite themselves, keeping their magnitude fainter than $V \gtrsim 7$ and $V_{\rm 1000km} \gtrsim 7$ is the single most efficient protection, as it makes them invisible to the eye and minimises the contamination. Satellites on lower orbit are also illuminated during a much shorter fraction of the night than those on higher orbit. At the stage of preparation of the observations, the level of contamination can be addressed statistically by pointing into the dark part of the constellations, as illustrated by Fig.~\ref{fig:skymap}. With a fairly significant effort, it is also possible to check whether an exposure about to start will be crossed by a satellite, by computing the position of all known satellites --provided that their orbital elements are know with suitable accuracy and precision, which is challenging for small field-of-views. 

An alternative is to equip the telescope with a camera that monitors $\sim10-15\degree$ around the observed field, detects incoming satellites, and either close the main shutter for the short time when the satellite passes through, or flag the data as possibly contaminated. \citet{sun23} describes such a system for the 4MOST fibre-fed spectroscopic survey. For each exposure, it will flag which of the 2400 fibres were affected by a satellite.

For imaging observations, data processing can detect and mask the trail. For spectroscopic observations, as discussed above, detecting the contamination can be challenging if the satellite's effective magnitude is close to that of the science object.
For surveys producing a large number of spectra with similar characteristics, the power of machine learning can be unleashed. \cite{bia+23} proposed such a system, based on convolutional network, to identify contaminated spectra. Based on a set of UVES (high-resolution echelle spectra) adjusted to the characteristics of the WEAVE stellar survey, they trained their system using artificial contamination (with a methodology similar, tough more sophisticated, to that described above). Their method also retrieve the astrophysical parameters of the stars, either after subtracting the contamination or in spite of the contamination. While they discuss issues and challenges, that paper is extremely promising for large surveys such as those of WEAVE and 4MOST. Indeed, while a majority of low level ($<5$\%) contamination are not detected, they report an impressive detection of 97.7\% true negative and 85.4\% true positive detection. \B{This method may be applied to single spectra, provided that the system has been trained on suitable (possibly simulated) data.}

For the general case, however, the best protection will be to take multiple observations, splitting the total exposure time in 2, 3 or more exposures. Fortunately, this technique is routinely used to detect and correct for cosmic rays.

The coordination among all stake holders provided by the IAU CPS is critical to share information among the community and with the satellite operators, reach common understanding of the issues, implement recommendations which can evolve into regulations.

\section{Conclusions}\label{sect:concl}

We studied how constellation satellites contaminate optical spectroscopy.  We used a pessimistic set of constellations totalling over 400\,000 satellites and simulated how often they crossed a $10\times1''$ slit during a 1000\,s exposure. The results depend on where the telescope points, but on average, about 10\% of the twilight exposures are affected by satellites, dropping to 0\% when the Sun is below 30$\degree$ under the horizon. The average over a whole night is about 2\%. These numbers are proportional to the slit length, the exposure time, and the number of satellites. 
The chance of a satellite crossing the observed object also varies with its apparent size.

Because of its fast apparent motion, the satellite spends a small fraction of a second in the slit, resulting in a contamination level equivalent to that of a (static) object of $M_{\rm eff}$ as defined in Eq.~\ref{eq:effmag1} and~\ref{eq:effmag2}. For constellation satellites, $M_{\rm eff}$ is in the range $\sim$14--22, overlapping with the magnitude range of the science targets. 
The contamination is negligible if the target is much brighter than $M_{\rm eff}$,  and obvious if the target is much fainter.

 We studied the intermediate case using a series of actual spectra, adding a scaled solar analogue spectra to simulate the effect of a satellite,  measured various astrophysical parameters and compared them with the non-contaminated values.  We found that the contamination affects the parameters at the percent level for high signal-to-noise spectra. This confirms that better data are more robust to contamination. We also noted that other satellites, such as bright ones (like ISS) or high ones (moving slower), can cause similar or worse contamination and that constellation satellite on transit orbit (between launch and operations) must also be kept at $V>7$. 

Overall, the fraction of contaminated spectra will be low, thanks to the small field of view of the \B{spectrographs's IFU(s), slit(s) or fibre(s)} and the fact that, due to their resolving power, many spectrographs are blind to constellation satellites, but contamination by satellites have the potential to be difficult to identify. While advanced data analysis tools will be able to flag some, the best protection against satellite contamination is the same that is used against asteroid and cosmic rays: split the total exposure time in two or more sub-exposures.

\begin{acknowledgements}
Based on observations collected at the European Organisation for Astronomical Research in the Southern Hemisphere under ESO programme(s) 60.A-9022(C), 189.A-0424(A),  096.D-0055(A), and 108.23MQ.001. 
\end{acknowledgements}

%-------------------------------------------------------------------
\bibliographystyle{aa} % style aa.bst
\bibliography{satConSpectro} % your references Yourfile.bib
\end{document}